# Topological metal behavior in GeBi$_2$Te$_4$ single crystals

A. Marcinkova[1], J. K. Wang[1], C. Slavonic[2], Andriy H. Nevidomskyy[1], K. F. Kelly[2],

Y. Filinchuk[3], and E. Morosan[1]

[1] Department of Physics and Astronomy, Rice University, Houston, Texas, 77005 USA

[2] Department of Electrical and Computer Engineering, Rice University, Houston, Texas, 77005 USA

[3] Institute of Condensed Matter and Nanosciences, Université Catolique de Louvain,
1348 Louvain-la-Neuve, Belgium

The metallic character of the GeBi$_2$Te$_4$ single crystals is probed using a combination of structural and physical properties measurements, together with density functional theory (DFT) calculations. The structural study shows distorted Ge coordination polyhedra, mainly of the Ge octahedra. This has a major impact on the band structure, resulting in bulk metallic behavior of GeBi$_2$Te$_4$, as indicated by DFT calculations. Such calculations place GeBi$_2$Te$_4$ in a class of a few known non-trivial topological metals, and explains why an observed Dirac point lies below the Fermi energy at about − 0.12 eV. A topological picture of GeBi$_2$Te$_4$ is confirmed by the observation of surface state modulations by scanning tunneling microscopy (STM).

## I. INTRODUCTION

Topological insulators (TIs) have recently emerged as a novel electronic state of quantum matter.[1-4] TIs have been rapidly recognized as materials exhibiting unique physical properties, such as Majorana fermions,[5] magnetic monopoles,[6] and with potential for applications in quantum computing.[7] However, the TI samples available today are invariably conducting in the bulk, and this bulk current always dominates the measured electrical transport. This poses the challenge of how to gain experimental control over the bulk and surface conductivity independently.

Theoretically, TIs are electronic materials that have a bulk band gap such as an ordinary insulator, but have gapless spin-polarized conducting edge states (in two-dimensional TIs) or surface states (in three-dimensional TIs), which are topologically protected. The topological protection means that the surface states are stable against local perturbations, such as impurities for example, and arises due to time-reversal symmetry.[8] Topological surface states usually show characteristic Dirac-like band dispersion, and exhibit spin-momentum locking. Evidence of surface states in TIs has been observed with surface sensitive probing techniques such as angle resolved photoemission spectroscopy (ARPES),[3,9] scanning tunneling microscopy (STM),[10,11] and, more recently, with transport measurements. A number of TIs, such as Bi$_2$Se$_3$ and Bi$_2$Te$_3$,[12-18] have been intensively studied. Recently, several pseudobinary compounds, such as Bi$_2$Se$_2$Te (BiTe-BiSe$_2$) or GeBi$_2$Te$_4$ (GeTe-Bi$_2$Te$_3$, GBT from now on), have been theoretically predicted to be three-dimensional TIs.[19,20] GBT belongs to the family of tetradymite-like layered pseudobinary compounds with a general chemical formula $A^{IV}B^{VI} - A_2^V B_3^{VI}$ ($A^{IV}$ = Ge, Sn, Pb; $A^{VI}$ = Bi, Sb; $B^{VI}$ = Te, Se).[21-24] Compounds with such a complex, many-layered structure (up to nine atomic layers per building block) had already been studied mainly due to their thermoelectric properties.[21-24] Very recently, GBT was claimed to be a 3D TI based on ARPES measurements,[25] albeit with a Dirac point lying below the Fermi energy (− 0.2 eV). This is in contrast with first principle calculations,[25,26] which place the Dirac point inside the band gap, and a resolution of this point is imperative.

In this paper, we investigate the surface states and bulk properties of GBT. The precise determination of the crystal structure and the atomic positions, performed using a high resolution synchrotron radiation X-ray source, reveals a small distortion of the Ge octahedra brought about by the reduction of the Bi-Te2 bond length. This distortion has a significant impact on the Fermi surface topology, and it results in the energy shift of the observed Dirac point below the Fermi level. The density functional theory (DFT) band structure calculations render GBT as a non-trivial topological metal, with the surface states characterized by a Dirac-like cone centered at ∼ − 0.12 eV, in agreement with previous ARPES measurements.[25] Moreover, STM data show a surface modulation similar to the one observed in other topological materials.

## II. EXPERIMENTAL METHODS

Single crystals of GBT were synthesized using a flux growth method. Elemental Ge (Alfa Aesar, 99.9999%), Bi (Alfa Aesar, 99.999%) and Te (Alfa Aesar, 99.99%) in an atomic ratio of 1:2:8 were packed in an alumina crucible and sealed in an evacuated quartz ampoule. The ampoule was heated up to 580°C, kept at that temperature for 2 hours, then slowly cooled down to 480°C, when the excess flux was decanted. The as-grown crystals were thin plates with typical dimensions of 7 x 5 x 0.3 mm$^3$.

High resolution synchrotron X-ray powder diffraction (SXPD) patterns were collected using the diffractometer at the Advanced Photon Source on beamline 11-BM.[27] Data at T = 100 K was collected using a Nitrogen gas cryostream. Two sample capillaries were measured using wavelength λ = 0.41 Å. Due to the presence of Bi and its high X-ray absorbance, the sample was diluted with amorphous SiO$_2$ powder in a molar ratio GBT:SiO$_2$ = 1:4. The sample was placed in a capton capillary and spun for better powder averaging. Data sets were collected between 2 ≤ 2θ ≤ 25° with a scan speed of 0.01°/s and binned with a step size of 0.001°. Analysis of the powder diffraction data was performed using the FullProf suite.[28] A pseudo-Voigt function was used to describe the peak shape for all data.

To further investigate the stoichiometry of GBT, X-ray photoelectron spectroscopy (XPS) was carried out using a PHI Quantera XPS scanning microprobe, with an Al-Kα scanning source. Zero field cooled (ZFC) DC magnetic susceptibility was measured using a commercial Quantum Design (QD) Magnetic Property Measurement System (MPMS). The electrical resistivity ρ with no applied magnetic field (H = 0) was measured in a standard four point geometry, using the resistance option of a Quantum Design Physical Property Measurement System (QD PPMS). The sample was cut into a bar-like shape, and four platinum wires were attached to the flat surface using Epo-Tek H20E silver epoxy, such that the current $i$ was confined to the $ab$ crystallographic plane. H = 0 heat capacity measurements were also carried out in the QD PPMS environment, using an adiabatic thermal relaxation technique. To determine the theoretical bulk and surface band structures of GBT, DFT band structure calculations were performed using the full-potential linearized augmented plane-wave method implemented in the WIEN2K package.[29] The calculations were then complemented by room-temperature scanning tunnelling microscopy (STM) measurements, performed in a RHK UHV-300 system with a base pressure of 10$^{-10}$ Torr. To produce fresh surfaces for imaging, the samples were cleaved *in situ* via leverage of a small rod epoxied to the top of the sample.

## III. RESULTS AND DISCUSSIONS

### A. Room Temperature Crystal Structure

Although $A^{IV}B^{VI} - A_2^V B_3^{VI}$ compounds have been widely studied for their thermoelectric properties, some controversy regarding the crystal structure and atomic positions still remains.[19-22, 30] The systems were found to have either an A- or B-site deficiency,[23] or antisite disorder.[22] Both have been reported previously[31, 32] and also observed recently by Okamoto et al.[26]

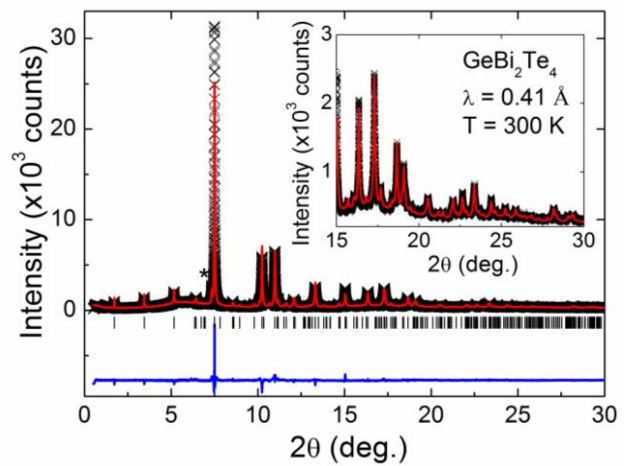

**Fig. 1 (color online) Rietveld refinement profile for the room temperature synchrotron X-ray powder diffraction data for GeBi$_2$Te$_4$. The difference between the measured data (red) and Rietveld fit (black) is shown as a blue line, the calculated Bragg positions are indicated by vertical markers. The secondary phase (asterisk) corresponds to 6.7 weight % of Bi.**

In the current work, an inspection of the synchrotron X-ray powder diffraction patterns of GBT confirmed the rhombohedral $R\bar{3}m$ structure, with lattice constants at room temperature $a = 4.3262(1)$ Å and $c = 41.3356(7)$ Å. These are in good agreement with previously reported room temperature values $a = 4.322(5)$ Å and $c = 41.127(2)$ Å.[23]

Several structural models were reported for GBT due to an occupational disorder on the metal sites. A model described by Karpinski et al.[22] consists of four atomic sites occupied by two or three elements as follows: (1) 3a (0,0,0) is occupied by Bi and Ge in a ratio 0.5:0.5; (2) 6c (0,0,0.4273) is shared by Bi:Ge:Te in a ratio 0.65:0.25:1; (3) 6c (0,0,0.1344) is occupied by Bi:Te in a ratio 0.97:0.03; and (4) is 6c (0,0,0.2903) with fractional occupancies 0.93:0.07 for Te:Bi. Another model by Agaev et al[30] is based on a fully ordered structure,

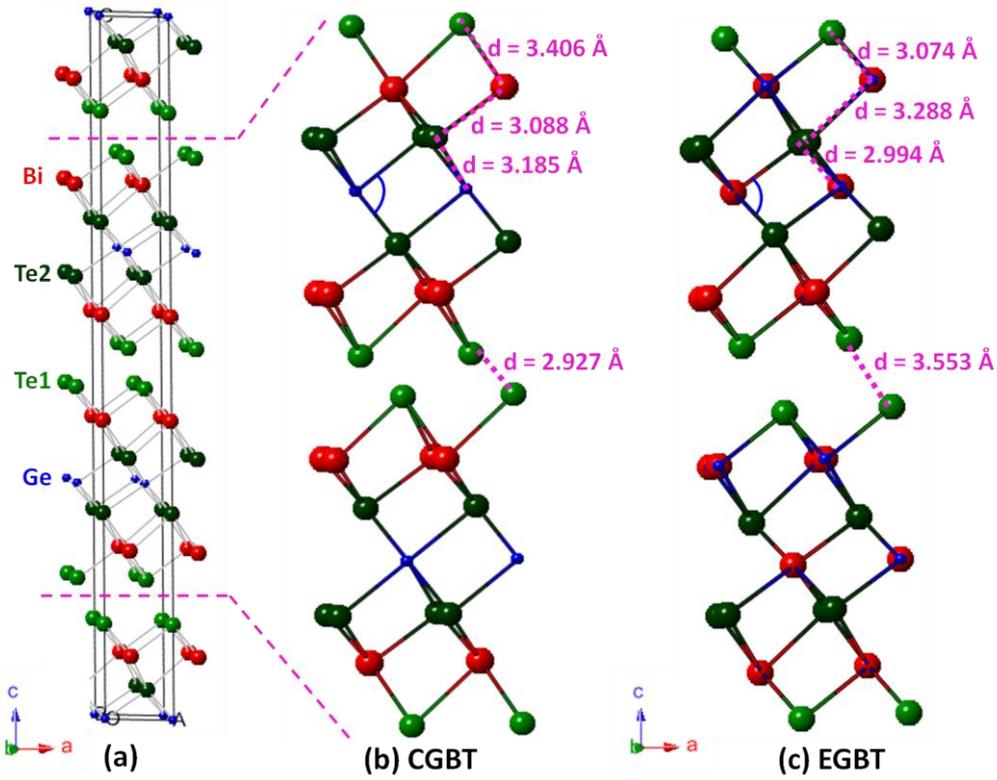

**Fig. 2** (a) Experimental room temperature crystal structure of GeBi$_2$Te$_4$ based on the Rietveld analysis of the room temperature synchrotron X-ray diffraction data. Details of the atomic arrangement in the (b) calculated and (c) experimental structural models for GeBi$_2$Te$_4$ (see text).

with (1) 3a (0,0,0) occupied by Ge; (2) 6c (0,0,0.4273) by Bi; (3) 6c (0,0,0.1344) by Te(1), and (4) 6c (0,0,0.2903) occupied by Te(2). A third model by Shelimova *et al.*[23] is a slightly modified version of the second model, in which a Ge-deficiency (2-3%) has been claimed. We propose a fourth model where the Ge/Bi mixing present only on the first site is induced as follows: (1) 3a (0,0,0) Ge:Bi in a ratio 0.5:0.5, while the other three sites remain fully ordered, (2) 6c (0,0,0.4273) Bi; (3) 6c (0,0,0.1344) Te(1), and (4) 6c (0,0,0.2903) Te(2).

We started the refinement with the fully ordered model by Agaev, which yielded a decent fit: $R_F$ = 11.6%, $R_p$ = 7.8%, $R_{wp}$ = 10.7%, $\chi_2$ = 7.8. The isotropic atomic displacement (ADP) for the 3a site turned negative, $B = 8\pi^2 U = -1.33(5)$ Å$^2$, while the Bi site got slightly too high: $B = 4.22(5)$ Å$^2$. The ADPs for Te varied in a reasonable 0.5-0.9 Å$^2$ range. This combination suggested Ge/Bi disorder on the first two sites, which was introduced with a single refined parameter between the two sites, keeping the total Ge:Bi ratio 1:2. This has improved the quality of the fit considerably, leaving nearly no notable differences between the experimental data and the calculation, except for the line shape of the strongest peak (Fig. 1). The final atomic positions, occupancies and ADPs are: (1) 3a (0,0,0) 0.509(2)Ge + 0.491(2)Bi, B = 2.21(7); (2) 6c (0,0, 0.42749(2)) 0.754(1)Bi + 0.246(1)Ge, B = 2.03(3); (3) 6c (0,0, 0.13405(2)) Te(1), $B$ = 1.07(3); and (4) 6c (0,0, 0.28990(3)) Te(2), B = 1.06(3). The very close and reasonable ADPs confirm disorder on Ge/Bi sites and no disorder on Te sites. The pattern shows 6.7 weight % of elemental Bi present, which should slightly decrease the Bi content on the Ge/Bi sites, thus bringing the ADPs close to those of Te atoms. The Bi impurity phase could be attributed to a residual surface flux, consistent with the Bi solution used for crystal growth. All together, this model is self-consistent and well parameterized, and is closest to the fourth model described in the previous paragraph. The final reliability indices are $R_F$ = 6.8%, $R_p$ = 5.8%, $R_{wp}$ = 7.2%, $\chi_2$ = 3.6.

## B. Calculated vs. Experimental Crystal Structure

The GBT structure contains building blocks consisting of septuplet atomic layers (Te1-Bi-Te2-Ge-Te2-Bi-Te1) as shown in Fig. 2a. The Bi-Te1 bonds had been found to be covalent with minor ionic character, while the Te1-Te1 bonds were stabilized by weak van der Waals forces.[20]

In order to study the topological nature and bulk character of GBT, we performed a structural optimization within DFT, using the experimental crystal lattice parameters and atomic positions as a starting point for the calculations. The resulting optimized crystal structure (*calculated GBT*, from now on CGBT), is shown in Fig. 2b, by comparison to the experimental one (*experimental GBT*, from now on EGBT) in Fig. 2c.

The calculated lattice parameters of CGBT $a = 4.36$ Å and $c = 41.3$ Å are very close to the experimental ones $a = 4.33$ Å and $c = 41.33$ Å determined for the EGBT. Although the cell constants do not change dramatically, the distance between the building blocks, *i.e.* between the septuplet atomic layer units, decreases abruptly from $d_{EGBT}$ (Te1-Te1) = 3.553 Å to $d_{CGBT}$ (Te1-Te1) = 2.994 Å. Such a change makes the Te1-Te1 bond not only covalent, but also the shortest bond in the calculated structure. The decreasing distance between the blocks in CGBT is caused by an elongation of the Te1-Bi bond $d$(Te1-Bi) from 3.074 Å (EGBT) to 3.406 Å (CGBT), and the Te2-Ge bond $d$(Te2-Ge) from 2.997 Å (EGBT) to 3.185 Å (CGBT). By contrast, the Bi-Te2 bond becomes shorter bond in the CGBT structure; $d$(Bi-Te2) = 3.088 Å, compared to the value in EGBT $d$(Bi-Te2) = 3.288 Å. The changing bonds result in the distortion of the Ge coordination polyhedra. In EGBT, the angle Te2-Ge/Bi-Te2 is equal to 87.8°, smaller than the ideal octahedral angle (90°). After the Ge-Te2 bonds expand and the Bi-Te2 bonds shrink (CGBT), the Te2-Ge/Bi-Te2 angle in CGBT becomes larger (94°). These structural differences between the calculated and experimentally measured structures will be shown below to have a significant impact on the band structure and Fermi surface topology of GBT. For completeness, the values of the Te1-Bi-Te1 angles are 94.3° (EGBT) and 96.1°.

All structural changes for both models, including the bond distances and the octahedral distortion, are shown in Fig. 2b, c.

### C. Magnetic and Electronic Properties

From the transport properties of topological materials it is hard to separate the surface state contribution from the bulk transport. Even a small conductivity from the imperfections in the bulk overwhelms the surface contribution. In order to determine the bulk contribution, a Landauer formula for resistance is used: $R_Q = h/(2e^2N)$, where $N$ is the number of edge channels, $e$ is the electron charge, and $h$ is Planck's constant.[33, 34] For a finite sample width, the conductance channels are bent at the edges of the samples. For each conductance channel intersecting the Fermi energy, one-dimensional channel, a so-called edge channel ($N$), is formed. In other words, $N$ corresponds to the trajectories of an electron moving along the edge of a sample, and is typically between 2 and 8. A bulk behaviour is metallic when the the number of channels N is N>>1, or in other words when the resistance $R_Q = h/2e^2$ is smaller than 25.8 kΩ.[34] This is indeed observed in GBT (Fig. 3a) where resistance is more than 3 orders of magnitude smaller than $R_Q$. The temperature dependence of the electrical resistance R for two GBT single crystals is shown in Fig. 3a.

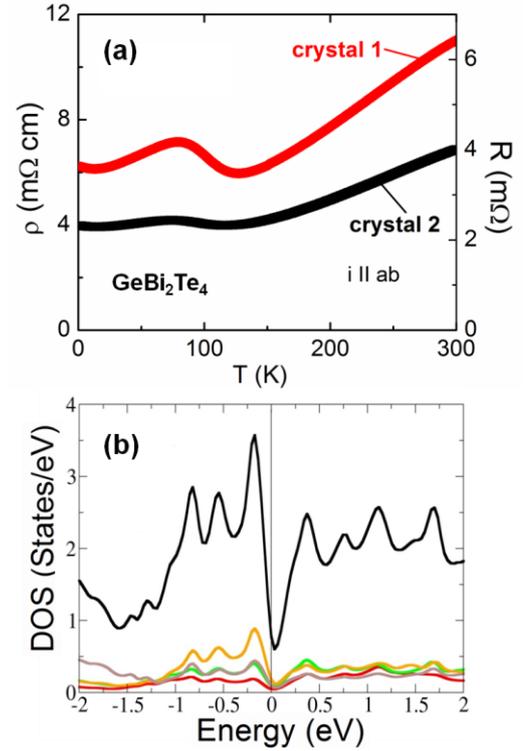

**Fig. 3 (a) Temperature dependence of the electrical resistivity of two GeBi$_2$Te$_4$ single crystals. (b) Calculated density of states of bulk of experimental GeBi$_2$Te$_4$ with contributions from all atoms**.

GBT exhibits metallic behaviour down to 100 K, consistent with the DFT calculations, which yields a non-vanishing density of states at the Fermi level (Fig. 3b).

The GBT resistivity has a similar temperature dependence to that of Bi$_2$Se$_3$.[12, 18, 35] However the absolute $\rho$ values, around 10 mΩcm, are several orders of magnitude smaller than those reported by Checkelsky *et al*.[35] This suggests a dominant metallic contribution of the bulk GBT to the transport properties. Furthermore Butch *et al*.[18] found, through a combination of Hall effect, transport and Schubnikov-de-Haas measurements on Bi$_2$Se$_3$, that there is no observable contribution from the surface states towards transport, *i.e.,* that the transport is entirely dominated by the bulk contribution.

A broad maximum is observed around $T \sim 80$ K in all measured samples, as illustrated by the measurements on two different crystals (Fig. 3a). Similar resistivity maxima have been observed for other TIs, for example in Bi$_2$Te$_3$.[36] As previously reported, the self-doping or vacancies on the Te sites in the Bi$_2$Te$_3$ TI lead to the insulating features in the resistivity.[36] Structural analysis of our synchrotron data yields fully occupied Te sites, but a very small deficiency on the Bi sites. Such Te excess (2-4%) would be enough to produce negative charge carriers (electrons), and be responsible for the observed resistivity maximum in GBT (Fig. 3a). Furthermore, varying the Te excess may be responsible, as was the case in previously reported TIs,[35] for the changes in the magnitude of the observed resistivity maximum.

To verify that the resistivity maximum is not associated with an intrinsic transition, specific heat and magnetization measurements were performed. No signatures of structural, electronic or magnetic phase transitions were observed down to 2 K.

### D. Band-Structure Calculations

In order to verify the topological nature of GBT, *ab initio* DFT calculations on the bulk material were performed, while artificially varying the strength of the spin-orbit coupling. The resulting electronic band structure is shown in Fig. 4. In order for the material to become a topological insulator, the band inversion must take place in some region of the Brillouin zone, usually at one of the high-symmetry points, but not in the entire zone. In this case, spin-orbit coupling leads to the gradual closing of the direct band gap at the $Z$ point (Fig. 4b-d) until the band inversion finally occurs when the strength of the spin-orbit coupling is close to 100% of its physical value (Fig. 4d). This suggests that GBT must have topological properties. However, unlike in $Bi_2Te_3$ and other "conventional" TIs, the calculations show that the indirect gap closes completely, revealing a non-vanishing density of states at the Fermi level as shown in Fig. 3b, meaning that GBT is a topological metal rather than an insulator. The experimentally measured metallic behavior of resistivity is consistent with this conclusion. It should be noted that the direct band gap remains open at the $\Gamma$ point in the bulk material (Fig. 4d). This should be contrasted with the situation in $Bi_2Te_3$, where the gap at the $\Gamma$ point closes for intermediate values of the spin-orbit coupling, whereas the gap at the $Z$ point always remains open.[9]

In order to unambiguously determine if the surface of the material hosts topological modes, calculations on a vacuum slab structure that contains 5 septuplet atomic layers (Te1-Bi-Te2-Ge-Te2-Bi-Te1, Fig. 5a) were performed. The Te1/Te1-terminated [001] surface was chosen with a cut across the Te1-Te1 covalent bonds (CGBT), as this is the most likely surface termination resulting in non-polar surfaces similar to $Bi_2Te_3$. Fig. 5 shows the calculated band structures of two slab structures: (b) the calculated structure (CGBT) that minimizes the DFT total energy, and (c) the crystal structure obtained from the refinement of SXPD data (EGBT). Surprisingly, the band structure of the optimized system in Fig. 5b differs significantly from the one in Fig. 5c, based on the experimental structure. The calculated structure in Fig. 5b displays a clear Dirac cone composed of surface states at the chemical potential, indicating that CGBT would be a topological insulator. In fact, the same conclusion had been reached theoretically in Ref. 25. This may imply that those conclusions were based solely on the CGBT structure, which is distinct from the distorted experimental structure (EGBT) found in this work. Our resistivity data however did not reveal any trace of insulating behavior (Fig. 3a), fully consistent with the metallic prediction for EGBT from Fig. 5c. In this case, the surface states form an electron pocket at the $\Gamma$ point, with the Dirac cone lowered to $-0.12$ eV below the Fermi level. Intriguingly, this is precisely the energy identified in the recent ARPES measurements.[25] The disorder on the Bi/Ge sites may lead to a more distorted Ge coordination polyhedra, which could lead to a shift of the Dirac point even lower in energy.

Further structural analysis is required to prove this point. Clearly, the experimental crystal structure (EGBT) used to calculate the bands in Fig. 5c appears to provide an accurate description of the GBT surface states, whereas the calculated structure in Fig. 5b, despite showing a clear Dirac cone at the Fermi level is actually incorrect. By comparing the two calculated band structures in Fig. 5b,c, it is readily apparent that the slight atomic displacement from the calculated structure results in the lowering of the electron bands at the $\Gamma$ point below the Fermi level. Since the overall system must remain charge-neutral, some bands away from the $\Gamma$ point must shift upward and form hole pockets, as is indeed the case in EGBT (Fig. 5c), which is consistent with the metallic behavior observed in our transport measurements. Note that the mechanism by which GBT becomes metallic is clearly different from that in $Bi_2Te_3$ or $Bi_2Se_3$ given that, in the latter cases, it is the self-doping or vacancies on Te/Se site that cause the shift in the chemical potential.[36]

### E. Scanning Tunneling Microscopy

Nanoscale characterization of the topological surface states in EGBT was carried out by room-temperature scanning tunneling microscopy (STM). An example of the typical images obtained by this procedure is shown in Fig. 6. Atomic resolution of the close packed surface of what is expected to be the Te1/Te1-terminated atomic plane (Fig. 6 insert) is shown to have an in-plane atomic spacing of

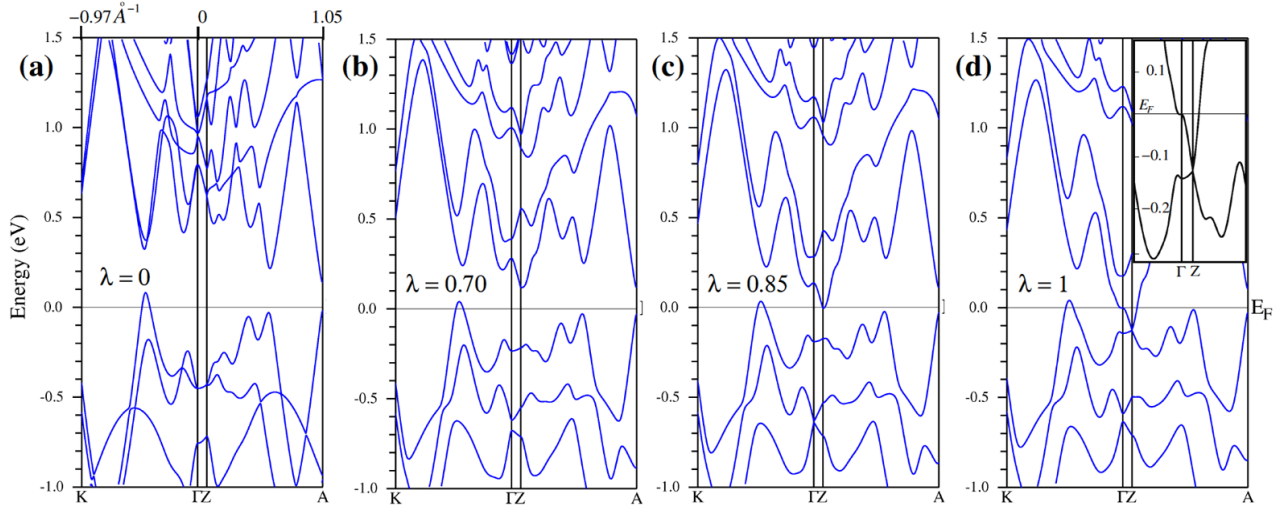

**Fig. 4** Band structure plots of bulk of experimental GeBi$_2$Te$_4$ for varying spin orbital coupling strength ($\lambda$): (a) in the absence of spin orbital coupling, (b) 70%, (c) 85%, (d) 100% of the physical spin orbital coupling. The inset in panel (d) depicts a zoomed-in band dispersion near the Fermi level (E$_F$), showing the Dirac cone of surface bands at about -0.12 eV below Fermi Energy (E$_F$).

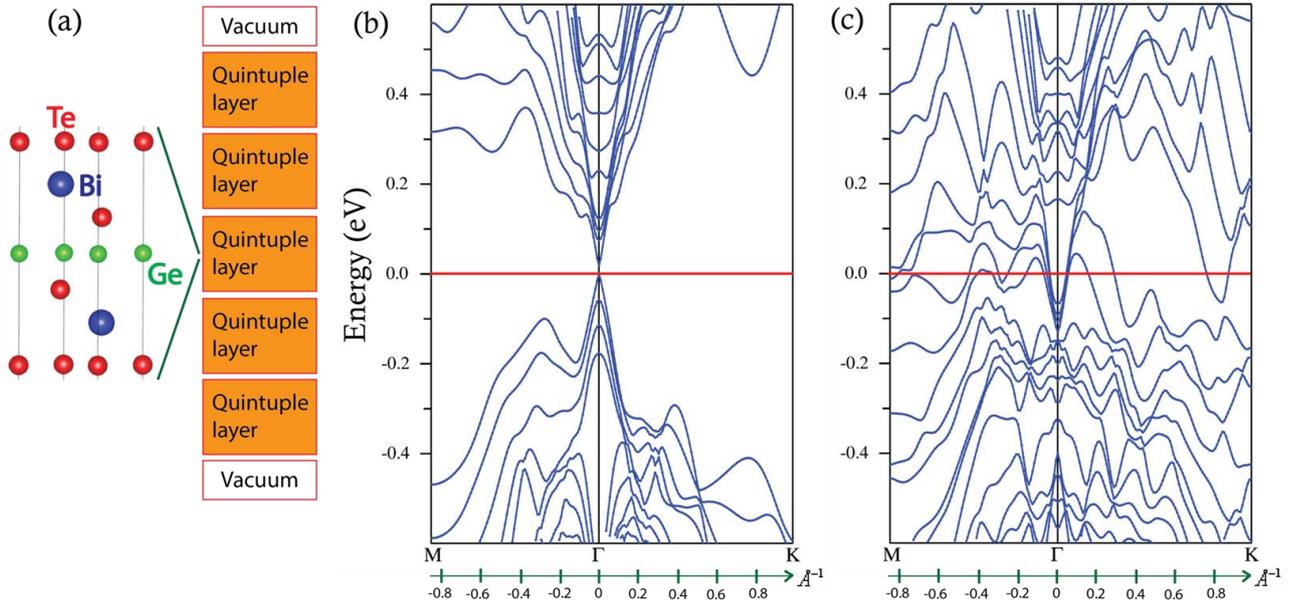

**Fig. 5** (a) Vacuum slab containing 5 atomic blocks, each comprised of 7 atomic layers, with the [001] surface terminated between the Te1-Te1 bond. (b, c) Calculated band structure of the slab geometry with atomic coordinates using (b) calculated and (c) experimental GeBi$_2$Te$_4$ structural models.

4.3 Å. While GBT is a layered compound in which cleaving should produce atomically flat planes similar to graphite, on larger-scale images a modulation is evident over the entire surface.

This is related to the interference of electrons in the surface state and appears because STM images are a convolution of both topographic and electronic structure. Images of electron scattering caused by perturbations, such as defects and step edges produce a modulation in the local density of states visible with STM.[13, 37-40] Here the modulation is electronic, not structural, in origin due to its sensitivity to the bias conditions, as shown in the Supplementary Material. Furthermore, the modulation resembles the states in topological materials[9] (and is unlike typical metallic surface states[39]) and represents allowed states around a constant energy slice of the Dirac cone. By calculating the two-dimensional fast Fourier transform (2D-FFT) of the real-space image, one gains insight into reciprocal space and the electronic structure of the material in the Brillouin zone. The image size of 75x75 nm was chosen to bring the atomic Bragg peaks to the edge of the 2D-FFT image (Fig. 7). As a reference, the 6 outer points (green circles) represent the atomic lattice points due to the close packed atomic structure. The longer wavelength states inside this hexagon represent scattering between topologically generated states.

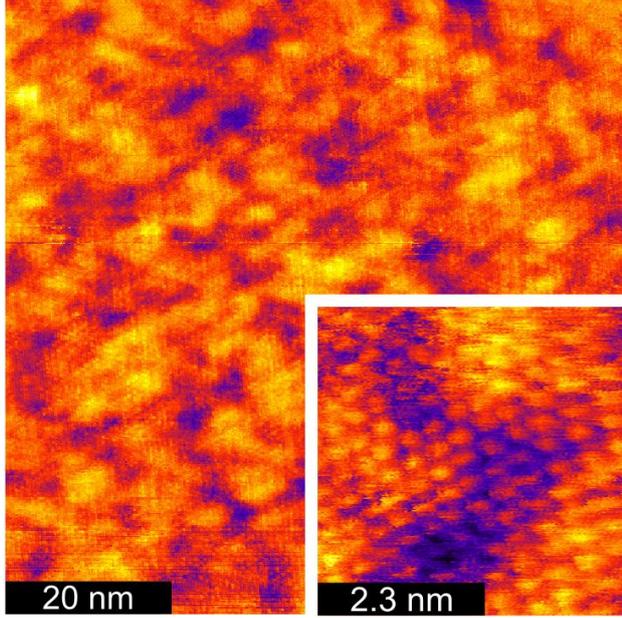

**Fig. 6** A 75x75 nm image of the $GeBi_2Te_4$ surface showing the local density of states modulation (see text). Inset; atomic resolution of the surface.

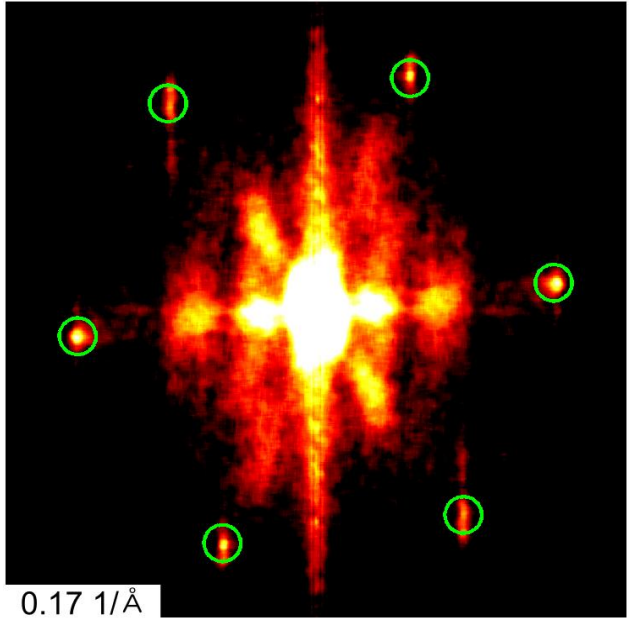

**Fig. 7** Calculated two-dimensional fast Fourier transform (2D-FFT) of the real-space. The image size of 75 nm was chosen to bring the atomic points to the edge of the 2D-FFT image. The six atomic lattice points due to the close packed structure are visible along the outer edge of the image (green circles), while the scattering signature is visible closer to the DC peak.

These additional inner peaks lie along the $K$ direction in line with the atomic points, with the peaks along the fast scan direction (horizontal) of the STM being most visible.

The non-circular contour of the point grouping is expected given the distance in energy between the sample bias voltage ($-35$ meV) and the Dirac point energy $E_D \approx -0.12$ eV, (from ARPES data and our DFT calculations), since the Dirac cone is known to become more circular as the energy approaches $E_D$ and to take on more hexagonal symmetry further away from it.[18, 25, 41-43] We see a peak at 0.136 Å$^{-1}$ in FT-STM reflecting the wavelength of the quasiparticle scattering at a sample bias of -35 mV. Since the topographic image represents the modulus squared of the wave-function, we calculate the wavenumber, $q_{surface}$, as 0.068 Å$^{-1}$.[47] Additional FT-STM images across both positive and negative biases will provide an analysis complimentary the ARPES measurements in mapping the electronic structure of GBT, and this is the subject of a future study.

## III. CONCLUSIONS

In summary, the combination of transport measurements, STM data and DFT calculations affirm that $GeBi_2Te_4$ is a metal which is, however, characterized by topologically protected surface states.

An inspection of the synchrotron data (Fig. 1) reveals Ge/Bi mixing and a distortion of the building blocks in EGBT compared to CGBT (Fig. 2c). Although GBT had been reported as a topological insulator, the present resistivity data suggests a metallic bulk character. Moreover, a Landauer theory[33] places GBT in a category of topological materials exhibiting metallic bulk behavior. DFT calculations reinforce this idea, as the DOS at the Fermi level is finite (Fig. 3b). The Dirac cone found from DFT at $E_D \approx -0.12$ eV is in good agreement with the existing ARPES studies, contrary to the fact that previous DFT calculations[26] on GBT which predicted the Dirac cone at $E_D \approx 0$ eV. Our DFT calculations using the experimental GBT structure (consisting of distorted Ge coordination polyhedra) yields the lower Dirac point energy, consistent with STM and ARPES data. STM data (Fig. 6) show surface states with a modulation similar to that observed in other topological materials[42] and distinct from non-topological regular metallic surface states.[41] More experimental studies (using different magnetic tips, doping into GBT) are necessary to explore the origins and character of such a modulation.

As demonstrated, the structural distortion of the Ge coordination polyhedra proves to be essential in converting the otherwise topological insulator to a non-trivial *topological metal*. This raises an intriguing question, as to whether the metallic bands in the bulk can disrupt the topological protection on the surface. It has been argued theoretically that, if the electron interactions were not too strong, the topological surface states would remain robust.[44, 45]

# APPENDIX

Figure A1 shows a series of scanning tunneling microscopy (STM) scans over -15 mV to -55 mV in 10 mV steps. The -35 mV image is the one included in the main paper. Highlighted are two distinct features (circle and square) on the surface to emphasize that the scans are all taken on the same region. There is a bit of drift due to piezoelectric hysteresis but other than that, the bias is the only variable changing between images. From this, we can safely say that the modulation is purely electronic and not topographic in nature.

Having established the electronic nature of this modulation, the only remaining discussion concerns its origin. What had been previously observed in STM topography at room temperature were features associated with charge density waves in transition metal dichalcogenides[46, 47], electron scattering in the presence of point defects[48-50] or impurities[51-53] in graphitic materials, or topological states.[9] All other quasiparticle scattering such as in superconductors and surfaces states of noble metals are only observable in topographic imaging at cryogenic temperatures. The present FTSTM images are distinct from those of either CDW or graphitic materials, but very much like the other cases of topological systems observed by STM. Therefore, it is reasonable to attribute the modulation observed in our STM data to topological states, which also agrees with the accompanying ARPES analysis and theory.

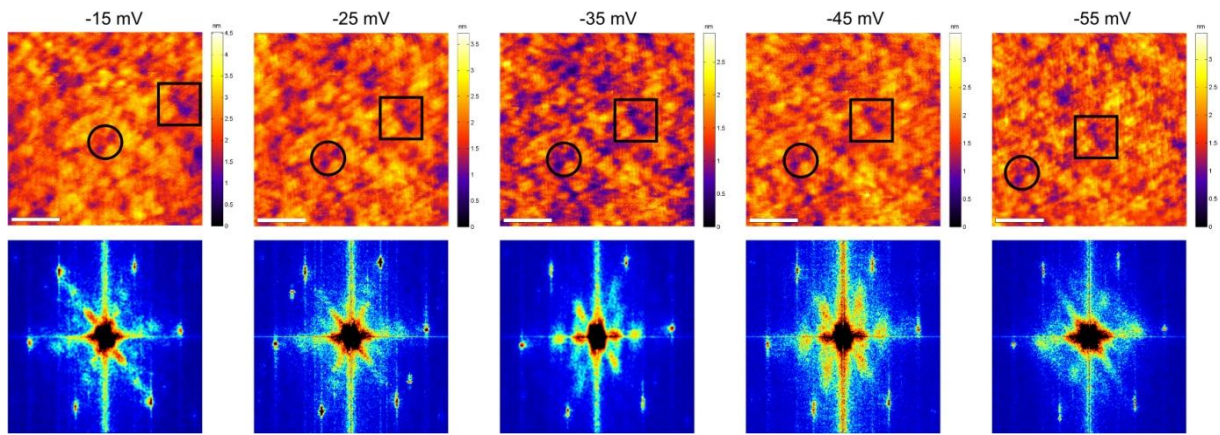

**Fig. A1:** A 75x75 nm image of the GeBi$_2$Te$_4$ surface showing the local density of states modulation (see text).

## ACKNOWLEDGEMENT

A.N. and A.M. thank Andrew Wray for fruitful discussions. E. M and A. M acknowledge support from the DOD PECASE award. K. F. K. acknowledges the Robert A. Welch Foundation (No. C-1605) for financial support. The authors thank Matthew R. Suchomel for the help with Synchrotron X-ray Powder Diffraction measurements. Use of the Advanced Photon Source at Argonne National Laboratory was supported by the U. S. Department of Energy, Office of Science, Office of Basic Energy Sciences, under contract No. DE-AC02-06CH11357.